\documentclass[conference]{IEEEtran}
\IEEEoverridecommandlockouts

\usepackage{cite}
\usepackage{amsmath,amssymb,amsfonts}
\usepackage{algorithmic}
\usepackage{graphicx}
\usepackage{textcomp}
\usepackage{xcolor}
\usepackage{comment}

\def\BibTeX{{\rm B\kern-.05em{\sc i\kern-.025em b}\kern-.08em
    T\kern-.1667em\lower.7ex\hbox{E}\kern-.125emX}}
\begin{document}

\title{A Distributed Energy Relay Framework for Sustainable Drone Missions
}

\author{
\IEEEauthorblockN{I-Ling Yen}
\IEEEauthorblockA{
\textit{Dept. Computer Science}\\
\textit{Univ. Texas at Dallas}\\
ilyen@utdallas.edu}
\and
\IEEEauthorblockN{Farokh Bastani}
\IEEEauthorblockA{
\textit{Dept. Computer Science}\\
\textit{Univ. Texas at Dallas}\\
farokh.bastani@utdallas.edu}
\and
\IEEEauthorblockN{Akeem Mohammed}
\IEEEauthorblockA{
\textit{Dept. Computer Science}\\
\textit{Univ. Texas at Dallas}\\
akeem.mohammed@utdallas.edu}
\and
\IEEEauthorblockN{Agastya Bose}
\IEEEauthorblockA{
\textit{Dept. Computer Science}\\
\textit{Univ. Texas at Dallas}\\
agastya@utdallas.edu}
}

\maketitle

\begin{abstract}
Limited battery endurance remains a major obstacle to persistent UAV missions. Existing energy-replenishment approaches typically rely on battery-swapping/charging infrastructure or on ground vehicles with accessible road networks -- assumptions that may not hold in infrastructure-less regions (ILRs) such as remote wilderness, mountainous terrain, and disaster-stricken areas. We formulate energy resupply in ILRs as a distinct logistics problem, in which energy itself must be projected into a region with no pre-existing infrastructure, and proposes a Distributed Energy Relay Framework that enables cooperative energy delivery using porter drones. The framework introduces a Transportation-by-Relay-and-Cache (TRC) mechanism, in which drones relay battery resources toward a mission location. Additional models are developed to characterize battery transportation capacity, relay amplification, energy-delivery capability, operational cost, and reliability. Based on these models, stations independently evaluate their delivery capability and submit bids to a mission coordinator, which selects an energy-delivery coalition that satisfies mission requirements while minimizing cost. Analytical evaluation illustrates the impact of battery characteristics and drone capabilities on energy projection and demonstrates the effectiveness of the proposed framework for sustaining UAV missions in ILRs.
\end{abstract}


\section{Introduction}
\label{sec:intro}
Unmanned aerial vehicles (UAVs) are increasingly employed in a wide range of applications, including package delivery, emergency response, environmental monitoring, infrastructure inspection, communication relay, and search-and-rescue operations. Many of these applications require persistent operation over extended periods of time, making energy availability a fundamental factor affecting mission success.
Despite significant progress in UAV technology, limited battery endurance remains a challenge in many drone missions. Most drones can remain airborne for less than an hour before requiring battery replacement or recharging.
Although the majority of drone-related works do not consider this energy-supply issue directly, a relatively substantial body of research has emerged over the last decade to address the drone energy problem specifically.

\subsection*{Related Work}

In \cite{Zeng2019}, a UAV flies along a continuous trajectory to provide wireless communication services to ground users whose locations are known a priori. It considers the UAV's energy consumption and optimizes its trajectory according to a fixed onboard energy budget. \cite{Masehian2024} seeks to ensure that a region is completely observed while minimizing overall mission completion time under a UAV energy constraint. It addresses the limitation of onboard energy but does not consider battery replacement or recharging.

Many drone missions require continuous execution for a significant amount of time. 
To address this, many works assume that drones originate from one or more base stations and periodically return to those stations for battery maintenance. In \cite{Dorling2017}, a multi-customer delivery problem is considered. It formulates an optimization problem that jointly determines the number of UAVs required, the assignment of customers to UAV trips, and the visiting sequence of customers within each trip, subject to payload capacity and energy constraints in each trip.
\cite{zhan2025balanced} considers a set of monitoring locations that must be revisited periodically by UAVs while meeting revisit constraints, with a focus on balancing the workload among UAVs. An improved $K$-means clustering algorithm is used to partition monitoring nodes among UAVs, each of which periodically returns to a fixed base station for battery replacement. \cite{arribas2023optimizing} addresses persistent service, where one UAV is required at each service location at all times: while a UAV is away replenishing energy, another must immediately replace it to maintain persistent service, and the objective is to determine the minimum fleet size and a resupply schedule that guarantees uninterrupted service at all locations. 
This general return-to-base strategy may not work for missions involving expansive infrastructure-less regions, where drones simply do not have sufficient energy to reach their mission locations, return to the station, or both.

A second line of works assume the availability of battery charging/replacement infrastructure. Several papers establish the concept of a ``skyway,'' in which drone and battery-management stations form a fully connected graph. In \cite{hamdi2021daas}, a delivery problem from a source station to a destination (customer location) is formulated along the skyway structure; the planner determines the best route from source to destination while avoiding weather-imposed risk, and for each route segment, a drone is selected to fly it based on remaining battery level and proximity.
\cite{mezni2025daas} considers the same skyway construction but, instead, considers station suitability, such as charger compatibility and historical performance, for route decisions. 
This skyway model depends on the stations themselves being accessible by ground or water vehicles; as an alternative to flying drones in a relay chain to a far-away destination, ground vehicles can transport the drones over most of the distance and let them fly only the last miles.

Some other works consider a similar static infrastructure for energy management. In \cite{Boucek2024}, multiple UAVs are each assigned a set of waypoints to visit, with static stations available for battery maintenance; the battery planner determines appropriate interruption points for battery replacement, and the scheduling problem is solved with an A*-based planner that searches for feasible replacement schedules given UAV battery state, mission progress, and station availability.
\cite{Tseng2017} and \cite{Alyassi2022} consider a similar problem, in which a single drone must visit a set of locations; charging stations are included as intermediate nodes in the planned route, and the optimization determines a feasible route that minimizes travel cost while satisfying all battery requirements.
\cite{Xu2020} considers a persistent surveillance problem in which each UAV is responsible for surveilling a region and is replenished through static charging stations, but the paper's focus is on station placement: UAVs are grouped by surveillance-task proximity and fuel feasibility, and for each group a min-max location problem places a static charging station within that group's region.
As with the skyway model, this line of energy replenish solutions heavily rely on infrastructure availability.

A third line of works employ unmanned ground vehicles (UGVs) for drone battery replenishment and/or transporting drones to the vicinity of mission locations, reducing infrastructure dependency to road or waterway network accessibility.
In \cite{Yu2019}, recharging resources include fixed stations and/or UGVs that act as mobile charging stations and move to rendezvous with the UAV; the UAV must visit a given set of target sites while minimizing total mission completion time, deciding the visit order, when to recharge, and where to land to recharge. \cite{shi2022riskaware} considers a similar setting but under energy-consumption uncertainty, formulating the rendezvous decision as a chance-constrained Markov decision process solved via linear programming. \cite{Tokekar2016} targets a precision-agriculture problem in which UAVs collect airborne agricultural information to support fertilizer decisions; a UGV transports the UAV and extends its operational range, and the system optimizes where to collect measurements.

\subsection*{Motivation}

Across all lines of works, one assumption is constant: some form of infrastructure is required to support the UAV operations. None of the literature considers what happens when no such infrastructure is available at all. 
With the increasing reliance on drones for special missions, we believe that in the near future drones will be deployed in infrastructure-less regions (ILRs) for many critical missions. For example, drones can be used to survey remote wilderness regions, perform rescue missions in mountainous terrain, or assist with communication relay in disaster-stricken areas.
In such ILRs, charging stations cannot be economically or safely deployed or maintained, conventional road/waterway networks are not readily available, and drones are the only viable platform for battery delivery. Indeed, these are precisely the scenarios in which drones are most useful and indispensable.

As a result, the problem of sustaining drone operations in an ILR becomes fundamentally an energy-logistics problem: mission success depends not only on the capabilities of the mission drones themselves, but on the system's ability to transport energy into the operational area.
This observation motivates a shift from traditional infrastructure-dependent drone support models toward energy-projection mechanisms that rely solely on cooperative drone operations. 
Rather than focusing exclusively on mission execution, some drones may be dedicated to transporting batteries that support other drones operating deeper within the ILR. Such cooperative behavior naturally leads to a relay-based logistics framework in which energy is progressively projected toward the mission location through multiple stages of drone-assisted transportation.

\subsection*{Overview}

In this paper, we propose a \textbf{Distributed Energy Relay Framework} for sustaining drone missions in infrastructure-less regions. The framework introduces a \textbf{Transportation-by-Relay-and-Cache (TRC)} mechanism in which porter drones cooperatively transport battery resources from stations surrounding (outside) the ILR to mission locations. The relay structure enables energy to be propagated over distances that exceed the flight endurance of individual drones, while strategically placed battery caches ensure that participating drones can safely return to their originating stations. The framework further supports heterogeneous drones and batteries and allows multiple stations to cooperatively contribute energy resources toward a common mission objective.

Building upon the TRC mechanism, we develop analytical models that quantify battery transportation capacity, relay amplification effects, energy-delivery capability, mission cost, and operational reliability. Each participating station independently evaluates its delivery capability and generates bids describing the energy, cost, and reliability of its proposed relay solution. A mission coordinator then selects an energy-delivery coalition that satisfies mission requirements while minimizing overall delivery cost. This formulation enables scalable coordination among geographically dispersed stations while maintaining operational flexibility.

\section{System Model and Problem Specification}
\label{sec:model}


Section~\ref{sec:intro} surveyed existing drone energy-replenishment strategies, including fixed charging and skyway networks, UGV-based recharging support, and continuous energy-transfer systems, and observed that all of them presuppose some form of infrastructures, whether a station, a transmitter, or a road or waterway network capable of supporting ground or water access. 
When such transportation modes remain available, the resulting problem concerns the deployment and management of these static and mobile infrastructures.

We consider a different scenario where a region cannot be accessed economically or safely by ground vehicles or maritime assets.
Such infrastructure-less region (ILR) may occur in highly remote wilderness regions, isolated mountainous terrain, post-disaster environments with severe access restrictions, or other locations where the deployment and maintenance of infrastructure become infeasible.
Instead of assuming that energy resources are available over the environment, one may explicitly consider how energy is transported into the operating region to support sustained missions.
This leads naturally to the concept of drone relay logistics, in which ``porter'' drones are responsible for transporting energy resources rather than mission execution. 
Mission feasibility is therefore determined not only by mission-execution capabilities but also by the collective ability of a drone fleet to project energy into an ILR environment.
Accordingly, this paper focuses on drone missions in ILR and develops a relay-based framework for extending mission sustainability through cooperative, autonomous energy delivery mechanisms.

\subsection{Problem Specification}

We consider an ILR that is surrounded by a set of drone stations that are accessible through road, waterway, or air transportation networks. Let $S = \{s_1,s_2,\ldots,...\}$ denote the surrounding $\lvert S \rvert$ stations that are close to the ILR and host drones, battery resources, and supporting facilities that may be used to support missions within the ILR. Each station $s_i$ is at location $L_i$ 

A mission request $M$ is represented by $M=(L^M,E^M,R^M)$, where $L^M$ is the mission location within the ILR, $E^M$ is the required energy to support the mission for the desired duration, and $R^M$ is the minimum acceptable mission reliability. Mission energy is measured in watt-hours (Wh). We consider heterogeneous batteries participating in the relay process, and all battery resources are converted into equivalent energy units.

For simplicity, we classify the drones as mission drones and porter drones. Mission drones travel to the mission location to provide the required mission services. We simplify the mission region to a single point $L^M$ in ILR and consider missions that may last for a significant duration.

Extensive prior works have investigated drone selection and service composition for mission execution
and we leave that aspect to existing studies.
This paper focuses on satisfying the energy needs at $L^M$ to support prolonged missions. Porter drones are responsible for battery transport and we consider a transportation-by-relay-and-cache (TRC) model for porter drones from multiple stations to deliver ``energy'' to meet the $E^M$ requirement of each mission. Here, relay is the structure formed by the porter drones for battery delivery and cache indicates that a drone can leave its semi-consumed battery at a certain location to be picked up and used later. It is assumed that the battery replacement can be accomplished by the drones autonomously. The detailed TRC mechanism is the focus of this paper.

There is a cost involved for each station to deliver batteries to $L^M$. Let $E_i$ denote the amount of energy delivered by $s_i$ and $C_i$ denote the transportation cost that $s_i$ incurs for the delivery. Cost model factors in the number of drones and the number of batteries required. Also, a drone may fail during a mission; thus, the more drones involved in a TRC solution, the lower the resulting reliability is expected to be. Based on its TRC solution, each station $s_i$ can compute $C_i$ and $R_i$, where $R_i$ denotes the reliability of the TRC solution of $s_i$.

The overall system goal can now be defined as a station selection problem as follows.
\[
\begin{aligned}
\min_{\mathbf{x}} \quad & \sum_{i=1}^{\lvert S \rvert} C_i x_i \\
\text{subject to} \quad
& \sum_{i=1}^{\lvert S \rvert} E_i x_i \geq E^{M}, 
\: \prod_{i=1}^{\lvert S \rvert} R_i^{x_i} \geq R^{M}, \\
& x_i \in \{0,1\}, \: i=1, \ldots, \lvert S \rvert.
\end{aligned}
\]
where $\mathbf{x}=(x_1,x_2,\ldots,x_{\lvert S \rvert})^T$ is the vector of binary decision variables, and $x_i=1$ if station $s_i$ is selected to execute its TRC, and $x_i=0$ otherwise.

Since each battery provides a non-divisible amount of energy, the total energy supplied by the selected stations should be allowed to exceed the required energy threshold. Therefore, the energy constraint is expressed as $\sum_{i=1}^{\lvert S \rvert} E_i x_i \geq E^{M}$, instead of an equality.

The TRC solutions are carried out by the porter drones. Mission drones may also require battery-transport support from porter drones to reach the destination before starting its mission execution using the delivered energy $E^M$. Since such support can be solved using the same TRC mechanism, we do not repeat the discussion and assume that mission drones have already arrived at $L^M$.

\subsection{Sketch of the Distributed Energy Relay Scheme}
Drone related missions are coordinated by a mission coordinator, who initiates missions either based on client requests or triggered by other missions. The mission is broadcast to the distributed stations. Each station derives its TRC solution and sends a bid to the coordinator. The coordinator uses a greedy algorithm to make selection decisions. The selected stations will execute their TRCs and deliver energy (i.e., batteries) to $L^M$.

To provide a comprehensive solution framework, we build the energy model for heterogeneous batteries, develop the relay scheme of TRC, and develop the cost and reliability models for each TRC solution. Finally, we design a greedy station-selection algorithm to accomplish the energy delivery goal.

\section{Battery and Energy Relay Models}
\label{sec:relay}
This section analyzes the TRC solution that each station can provide and the amount of energy that can be delivered to the mission location. Global coordination and decision making across multiple stations will be discussed in the next section.

\subsection{Battery Energy and Weight}
A battery type $b$ is characterized by its series cell count $S_b$ and capacity $C_b$ (Ah). For lithium-polymer (LiPo) batteries, each cell has a nominal voltage of approximately 3.7 V. Therefore, the nominal energy stored in battery $b$, measured in watt-hours (Wh), is
$E_b = 3.7S_bC_b$.
For example, a 3S battery with a capacity of 2.2 Ah stores
$E_b = 3.7 \times 3 \times 2.2 = 24.42\;\mathrm{Wh}$.

The weight of a battery can be modeled by two components:
(1) a fixed packaging overhead associated with connectors, balancing leads, casing, and protection circuitry; and
(2) the weight of the electrochemical cells.
Let $\rho$ denote the battery energy density (Wh/kg). The battery weight is modeled as
$W_b = W_0 + \frac{E_b}{\rho}$,
where $W_b$ is the battery weight, $W_0$ is the fixed packaging overhead, and $\rho$ is the battery energy density.

Together we get
\[
W_b = W_0 + \frac{3.7S_bC_b}{\rho}.
\]
In this work, $\rho = 180\;\mathrm{Wh/kg}$ and $W_0 = 0.02\;\mathrm{kg}$ are adopted as representative values, yielding
$W_b = 0.02 + \frac{3.7S_bC_b}{180}$.

\subsubsection{Drone Battery Transportation Capacity}
Let $P_d$ denote the payload capacity of drone type $d$. The maximum number of batteries of type $b$ that can be transported by the drone is
\begin{equation}
K_{d,b}=\left\lfloor \frac{P_d}{W_b} \right\rfloor.
\label{eq:carry_capacity}
\end{equation}
$K_{d,b}$ denotes the battery transportation capacity of a porter drone. Table~\ref{tab:battery_types} gives some data of representative batteries in the market and their corresponding energy to weight ratio.

\begin{table}[t]
\centering
\caption{Representative Commercial LiPo Batteries}
\label{tab:battery_types}
\begin{tabular}{lcccccc}
\hline
Battery & \#Cells & Capacity & Energy & Weight & Ratio \\
Model & ($S_b$) & (Ah) & (Wh) & (kg) & (Wh/kg)\\
\hline
Tattu 2200mAh 3S & 3 & 2.2 & 24.42 & 0.180 & 135.67 \\
Tattu 5000mAh 3S & 3 & 5.0 & 55.50 & 0.328 & 169.21 \\
Tattu 10000mAh 3S & 3 & 10.0 & 111.00 & 0.637 & 174.25 \\
Tattu 5000mAh 6S & 6 & 5.0 & 111.00 & 0.645 & 172.09 \\
\hline
\end{tabular}
\end{table}
Batteries with higher energy have a slightly higher energy to weight ratio. However, the real pros and cons of which batteries to transport depend on the transportation capacity of the drone. Unlike fuel in the jeep problem, the number of batteries a drone can carry is an integer value and, hence, the battery weight has a sensitive impact.

\subsection{Transportation-by-Relay-and-Cache}
We develop the TRC model for energy delivery by drones. In general, a drone can fly for 20--40 minutes on a single battery charge, covering a distance of approximately 10--20 miles. If the ILR spans a much larger area and the mission location is located near its center, multiple drones must collaborate in a relay-chain manner to successfully reach $L^M$. The central idea is to establish a sequence of relay locations through which battery resources are progressively transported.

\subsection{Drone Return Considerations}
In practical deployments, porter drones must return to their originating stations after completing their delivery missions. Let $R_{d,b}$ denote the effective relay range of drone type $d$ using battery type $b$ for propulsion, which is the maximum distance this drone can reach and return while maintaining a reasonable battery reserve. Generally, a drone battery remains healthy if its state of charge does not fall below approximately 20--25\%. Therefore, $2R_{d,b}$ represents the approximate round-trip distance that the drone can safely travel while maintaining the desired battery reserve. Instead of considering one-way endurance, $R_{d,b}$ incorporates the energy reserve required for safe return without damaging the battery.

For clear illustration, we first use the case in which a porter drone carries only one additional battery (referred to as the relay battery) besides its own propulsion battery. We then extend the formal derivation to a generalized payload capacity.

Suppose a porter drone $P$ departs from a relay point carrying its own propulsion battery and one relay battery. After flying one relay segment of length $R_{d,b}$, $P$ arrives at the next relay point with a partially depleted propulsion battery. If the mission terminates at that relay point, the remaining energy reserve is sufficient for the return trip. If the mission continues farther downstream, allowing $P$ to continue flying with its partially depleted battery is undesirable because the battery being transported would eventually be consumed by drone $P$ for its return, defeating the purpose of extending battery resources deeper into the relay chain.

To overcome this, a battery-refresh procedure is performed at every relay point. Upon arrival at a relay point, drone $P$ stores its partially discharged propulsion battery at the current relay location and replaces it with the fully charged battery being transported. The stored battery becomes a \emph{battery cache}. Drone $P$ is therefore restored to a fully operational state and can continue to the next relay segment while transporting another battery.

When drone $P$ later completes its delivery mission and begins returning toward its home station, it retrieves the battery cache that was previously left at each relay point. These cached batteries provide the energy required for the return journey on a segment-by-segment basis. The process repeats recursively until drone $P$ reaches its original station.

Under this operational model, battery caches guarantee the return of all porter drones while preserving the ability to project energy deeper into the mission region. Empty batteries generated during the process are assumed to remain at relay locations for subsequent collection and recycling. Battery-cleanup logistics are outside the scope of this paper and can be accomplished in a manner similar to battery delivery.

\subsubsection{Relay Chain}
For clarity, we first consider the case in which each porter drone can carry only one additional battery besides its own propulsion battery.

Consider a group of drones $\{P_1,P_2,\ldots\}$ from station $s_i$ that collaboratively deliver a single battery to $L^M$.
If the distance from $L_i$ to $L^M$ is equal to one relay segment $R_{d,b}$, then a single drone can deliver the battery and safely return.

If the distance is $2R_{d,b}$, two drones are required. Drones $P_1$ and $P_2$ travel together to the first relay point, located $R_{d,b}$ away from station $s_i$, carrying two relay batteries. At the relay point, $P_1$ returns toward the station and leaves its relay battery behind. Drone $P_2$ replaces its partially depleted propulsion battery with the relay battery left by $P_1$ and continues toward $L^M$ while carrying its original relay battery.

If the distance is $3R_{d,b}$, four drones are required. The four drones travel together to the first relay point carrying four relay batteries. Two drones then return toward the station and leave their relay batteries behind at the relay point. The remaining two drones replace their partially depleted propulsion batteries with these two relay batteries and continue carrying their own relay batteries. At this point, the two remaining drones are in the same state as the $2R_{d,b}$ case, except that their starting location is now the first relay point. The subsequent relay procedure proceeds recursively.

\subsubsection{Amplification Factor and Drone/Battery Requirements}

For the example case discussed above, the relay process forms a binary-tree structure with an amplification factor of two.
We now generalize the relay process by considering a drone of type $d$ that can carry $K_{d,b}$ batteries of type $b$.

Let ${dist}_i$ denote the straight-line distance from station $s_i$ to the mission location $L^M$. Since the straight-line path is the shortest, it is assumed to be the relay path followed by the porter drones. For drone type $d$ using battery type $b$, the relay depth, i.e., the number of relay segments required along this route, is defined as
\begin{equation}
 h_{i,d,b}
 =
 \left\lceil
 \frac{{dist}_i}{R_{d,b}}
 \right\rceil.
\label{eq:relay_depth}
\end{equation}

Let $K_{d,b}$ be the number of type-$b$ batteries that a drone of type $d$ can carry.
Suppose that $K_{d,b}$ batteries must be forwarded from relay level $r_1$ to the next relay level $r_2$ by a single drone. An additional battery is required at $r_1$ to support the relay transition, i.e., for the downstream drone to replace its partially depleted propulsion battery. Therefore, a total of $K_{d,b}+1$ batteries must be available at $r_1$ in order to forward $K_{d,b}$ batteries to $r_2$. This introduces an expansion factor of
\begin{equation}
\frac{K_{d,b}+1}{K_{d,b}},
\end{equation}
which represents the ratio between the batteries required at the current relay level and the batteries that can be forwarded to the next relay level.

For a mission that requires only one relay segment ($h_{i,d,b}=1$), no intermediate relay operation is needed and hence no amplification occurs. Amplification is introduced only when batteries must be forwarded through intermediate relay levels. Since there are $h_{i,d,b}-1$ relay transitions along a path of depth $h_{i,d,b}$, the overall amplification factor is approximated by
\begin{equation}
G_{i,d,b} = \left(\frac{K_{d,b}+1}{K_{d,b}}\right)^{h_{i,d,b}-1}.
\label{eq:relay_growth}
\end{equation}

The above expression ignores edge effects associated with the final relay level and integer rounding effects caused by discrete battery counts. The amplification factor represents the growth in logistics resources required to project energy across multiple relay levels while simultaneously guaranteeing the return of all porter drones.

More generally, suppose station $s_i$ must deliver $K_i$ batteries to the mission location through a relay chain of depth $h_{i,d,b}$. Due to relay amplification, the number of batteries that must be committed at the source station is 
\begin{equation}
N_{i,d,b} = \left(K_i \frac{K_{d,b}+1}{K_{d,b}}\right)^{h_{i,d,b}-1}
\label{eq:required_batteries_integer}
\end{equation}
where $N_{i,d,b}$ is the total number of batteries of type $b$ required at station $s_i$ to successfully deliver $K_i$ batteries to $L^M$.

The relay process also requires sufficient porter drones to transport these batteries through successive relay levels. Let ${ND}_{i,d,b}$ denote the total number of type-$d$ drones that must be committed by station $s_i$ for the corresponding relay operation for battery type $b$. Since each drone can transport at most $K_{d,b}$ type $b$ batteries during a relay flight, the number of drones required is proportional to the total battery inventory that must be conveyed through the relay network. Neglecting the small inefficiency caused by the final partially loaded drone, the required number of drones is approximated by
\begin{equation}
{ND}_{i,d,b} = \left\lceil \frac{N_{i,d,b}}{K_{d,b}} \right\rceil
=
\left\lceil \frac{\left\lceil K_i G_{i,d,b}\right\rceil}{K_{d,b}} \right\rceil.
\label{eq:required_drones_integer}
\end{equation}

Note that when we compute the required number of drones, ${ND}_{i,d,b}$, we do not account for the repeated use of a smaller set of drones during the energy delivery mission. 
Since ${ND}_{i,d,b}$ can be used to compute the cost and reliability of the relay mission correctly, we use it directly without further considering the trade-off between mission completion time and the number of drones used.


\section{Distributed Energy Relay Coordination}
\label{sec:coord}
When a station receives a mission request
$M=(L^M,E^M,R^M)$,
it independently evaluates its delivery capability using the transportation-by-relay-and-cache (TRC) model developed in the previous section. Specifically, the station determines the amount of mission energy that can be projected to the destination, the batteries and porter drones required to support the delivery, the associated delivery cost, and the expected delivery reliability. The stations then submit bids to a mission coordinator, which selects an energy-delivery coalition that satisfies mission requirements while minimizing the overall cost.

\subsection{Local Relay Computation}

Upon receiving a mission request, each station independently evaluates its local resources and derives one or more feasible relay configurations.

Let $D_{i,d}$ denote the number of available porter drones of type $d$ at station $s_i$, and let $B_{i,b}$ denote the number of available batteries of type $b$ at station $s_i$.

In principle, a station may host multiple drone types and multiple battery types. Furthermore, different drone types may be capable of transporting different battery types, and a drone may potentially transport a mixture of battery types during a relay operation. Determining the optimal assignment between heterogeneous drones and heterogeneous batteries therefore introduces an additional resource-allocation problem involving all feasible drone-battery combinations.

To focus on the TSC mechanism, we adopt a simplified model in this paper. A drone of type $d$ is assumed to operate with a predetermined battery type $b$. That is, each porter drone transports only batteries of its assigned battery type, and all relay calculations are performed for a fixed $(d,b)$ pair. The same applies to the propulsion batteries. More general problem involving multiple battery types per drone, multiple drone-battery assignments, and cross-type transportation decisions is left for future investigation.

For each relay configuration $(d,b)$, station $s_i$ computes the transportation capacity $K_{d,b}$, relay depth $h_{i,d,b}$, relay amplification factor $G_{i,d,b}$, and maximum deliverable battery count

\begin{equation}
K_{i,d,b}^{\max}
=
\left\lfloor
\frac{
\min
\left(
B_{i,b},
D_{i,d}K_{d,b}
\right)
}
{G_{i,d,b}}
\right\rfloor .
\label{eq:max_deliverable_batteries}
\end{equation}

The corresponding deliverable energy is

\begin{equation}
E_{i,d,b}
=
K_{i,d,b}^{\max}E_b.
\label{eq:max_deliverable_energy}
\end{equation}

The relay computation also determines the battery inventory required at the source station,

\begin{equation}
N_{i,d,b}
=
\left\lceil
K_{i,d,b}^{\max}
G_{i,d,b}
\right\rceil ,
\label{eq:required_inventory}
\end{equation}

and the number of porter drones required,

\begin{equation}
{ND}_{i,d,b}
=
\left\lceil
\frac{N_{i,d,b}}
{K_{d,b}}
\right\rceil .
\label{eq:required_porters}
\end{equation}

The quantities $(E_{i,d,b},N_{i,d,b},{ND}_{i,d,b})$ collectively characterize the delivery capability of station $s_i$ under relay configuration $(d,b)$ and form the basis for subsequent cost and reliability calculations.

\subsection{Cost Model}

The relay operation consumes both battery resources and drone resources. Therefore, the delivery cost is modeled as a combination of battery-utilization cost and drone-utilization cost.

Let $c_b$ denote the effective cost associated with allocating one battery of type $b$ to the relay operation. Similarly, let $c_d$ denote the effective utilization cost associated with one porter drone of type $d$.
For relay configuration $(d,b)$ at station $s_i$, the delivery cost is defined as

\begin{equation}
C_{i,d,b}
=
c_bN_{i,d,b}
+
c_d{ND}_{i,d,b}.
\label{eq:cost}
\end{equation}

The battery component captures the inventory committed to the relay process, while the drone component reflects the fleet resources required to sustain the relay chain.

\subsection{Reliability Model}

Let $\mathrm{MTTF}_d$ denote the mean time to failure of drone type $d$.
Assuming an exponential failure model, the reliability function of drone type $d$ is
\begin{equation}
r_d(t)
=
e^{-t/\mathrm{MTTF}_d},
\end{equation}
where $t$ denotes the mission duration.

We normalize one relay segment flying time as one unit time. Note that the drones fly different numbers of segments and we need to derive the expected number of segments flown by all the porter drones to properly derive reliability.

At each relay level, forwarding $K_{d,b}$ batteries to the next level requires $K_{d,b}+1$ batteries at the current level. Thus, with probability $p = (K_{d,b})/(K_{d,b}+1)$, a drone continues past a given relay level, and the probability for the drone to hand off its battery and turn back at this level is $1-p = 1/(K_{d,b}+1)$. 
The number of levels a drone advances before handing off, $k$, therefore satisfies $\Pr[k>j]=p^{j}$ for $j=0,1,\dots,h_{i,d,b}-1$, and its expectation follows the standard truncated-geometric identity $\mathbb{E}[k]=\sum_{j=0}^{h_{i,d,b}-1}\Pr[k>j]$, which yields 
\begin{equation} 
\begin{split} 
\bar{\tau}_{i,d,b} &= \mathbb{E}[k] 
= \sum_{j=0}^{h_{i,d,b}-1} p^{\,j} 
= \frac{1-p^{\,h_{i,d,b}}}{1-p} \\ 
&= \left(K_{d,b}+1\right)\left(1-p^{\,h_{i,d,b}}\right). 
\end{split} 
\label{eq:avg_exposure} 
\end{equation}



The relay operation requires ${ND}_{i,d,b}$ porter drones. Assuming failures among drones are statistically independent, the probability that all participating drones successfully complete the relay operation is
\begin{equation}
R_{i,d,b} = \left( e^{-\bar{\tau}_{i,d,b}/\mathrm{MTTF}_d} \right)^{ {ND}_{i,d,b}}
= e^{-\dfrac{{ND}_{i,d,b}\,\bar{\tau}_{i,d,b}} {\mathrm{MTTF}_d} }.
\label{eq:reliability}
\end{equation}

Note that only outbound flight contributes to $\bar{\tau}_{i,d,b}$, drone failures occurring on the return leg -- after batteries have already been delivered or handed off -- do not reduce $R_{i,d,b}$, because it does not contribute to the energy-delivery mission itself.

\subsection{Distributed Bidding}

After completing local relay computation, station $s_i$ generates a set of feasible bids
\[
\mathcal{B}_i=
\left\{
(E_{i,d,b},C_{i,d,b},R_{i,d,b})
\mid
(d,b)\in\mathcal{F}_i
\right\},
\]
where $\mathcal{F}_i$ denotes the set of feasible relay configurations available at station $s_i$.

Each station transmits its bid set to the mission coordinator. The coordinator performs global bid selection and determines which relay configuration, if any, should be activated at each station.

\subsection{Mission Coordination Problem}
We refine the definition for the decision variable $x_i$ give in Section \ref{sec:model} to the selection of drone and battery type pairs $(d,b)$, i.e. $x_{i,d,b}$.
The mission coordinator solves
\begin{equation}
\min
\sum_{i=1}^{\lvert S \rvert}
\sum_{(d,b)\in\mathcal{F}_i}
C_{i,d,b}x_{i,d,b}
\label{eq:objective}
\end{equation}

subject to

\begin{equation}
\sum_{i=1}^{\lvert S \rvert} \sum_{(d,b)\in\mathcal{F}_i} E_{i,d,b}x_{i,d,b} \ge E^{M} \: \land
\label{eq:energy_constraint}
\end{equation}

\begin{equation}
\prod_{i=1}^{\lvert S \rvert} \prod_{(d,b)\in\mathcal{F}_i} R_{i,d,b}^{x_{i,d,b}} \ge R^{M} \: \land
\label{eq:reliability_constraint}
\end{equation}

\begin{equation}
\sum_{(d,b)\in\mathcal{F}_i} x_{i,d,b} \le 1, \: \forall i.
\label{eq:one_bid_per_station}
\end{equation}



The resulting formulation is a multiple-choice 0--1 integer optimization problem.

\subsection{Greedy Energy Delivery Coordination}

Although the integer optimization formulation yields an optimal solution, rapid mission response may be required in large-scale deployments. We therefore adopt a lightweight greedy heuristic.
Each feasible bid computes a utility metric
\begin{equation}
U_{i,d,b}
=
\frac{C_{i,d,b}}
{E_{i,d,b}}.
\label{eq:unit_cost}
\end{equation}
representing the delivery cost per unit of mission energy.

All feasible bids are sorted in ascending order of $U_{i,d,b}$. The coordinator iteratively selects bids
until both the mission-energy and mission-reliability requirements are satisfied.



\section{Analytical Evaluation}
\label{sec:eval}
We analytically evaluate our transportation-by-relay-and-cache (TRC) framework using the models developed in Sections \ref{sec:relay} and \ref{sec:coord}. The objective is to investigate how mission distance, battery selection, and porter-drone capabilities influence the amount of energy that can be projected into an infrastructure-less region (ILR).

Unless otherwise stated, the battery energy density is set to $\rho=180$ Wh/kg and the battery packaging overhead is set to $W_0=0.02$ kg. The station is assumed to contain $D_{i,d}=20$ porter drones and $B_{i,b}=100$ batteries of the evaluated drone and battery types. Mission distance varies from 10 km to 60 km. Three representative porter-drone classes are considered:

\begin{itemize}
\item Light porter drone: payload capacity 0.5 kg and relay range 5 km.
\item Medium porter drone: payload capacity 2 kg and relay range 10 km.
\item Heavy porter drone: payload capacity 8 kg and relay range 15 km.
\end{itemize}

The battery models listed in Table~\ref{tab:battery_types} are used throughout the evaluation.




A


\subsection{Battery-Type Comparison}

\begin{figure}[t]
\centering
\includegraphics[width=0.9\linewidth]{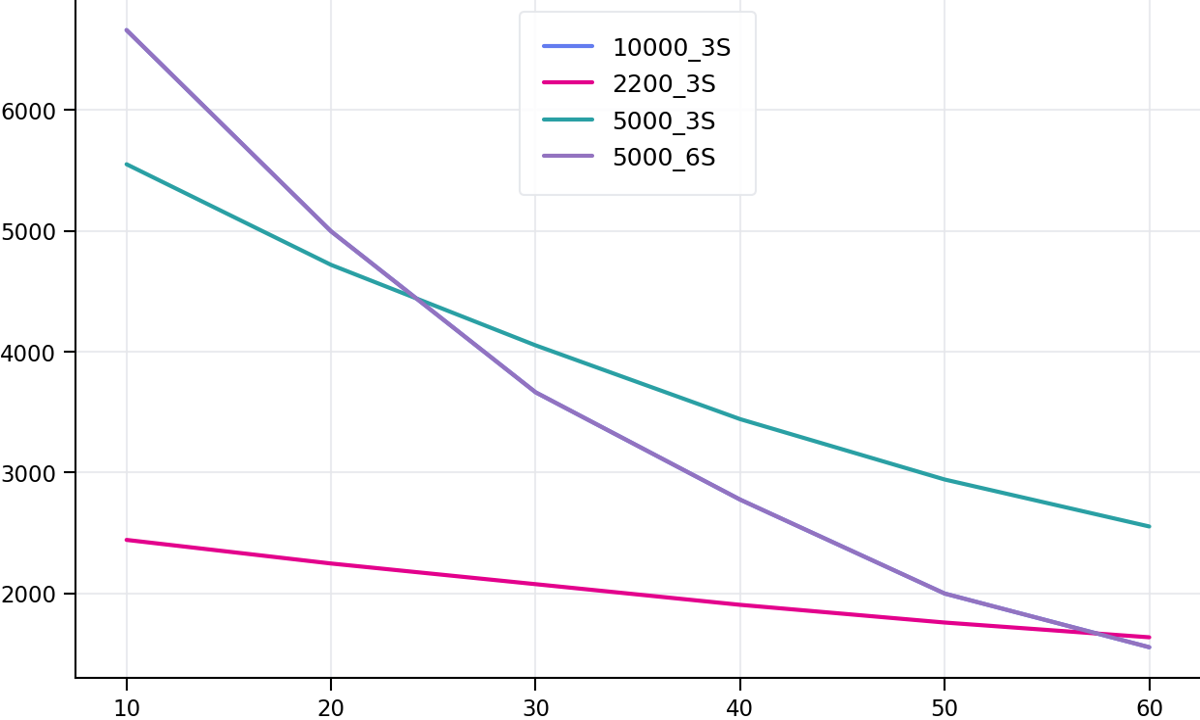}
\caption{Deliverable energy versus mission distance for different battery types using a medium porter drone.}
\label{fig:battery_comparison}
\end{figure}

We investigate how battery selection affects energy deliver efficiency. Figure~\ref{fig:battery_comparison} shows the deliverable energy as a function of mission distance for different battery types using the medium porter drone.

The medium porter drone has a payload capacity of 2 kg and a relay range of 10 km. Using the battery-weight model developed in Section \ref{sec:relay}, the resulting transportation capacities are approximately 12, 6, 3, and 3 batteries for the 2200 mAh 3S, 5000 mAh 3S, 10000 mAh 3S, and 5000 mAh 6S batteries, respectively. Thus, smaller batteries allow substantially larger transportation multiplicity while larger batteries provide significantly more delivered energy per battery.

Figure~\ref{fig:battery_comparison} shows that larger battery models generally achieve substantially higher deliverable energy than smaller battery models across the entire distance range. In particular, the 10000 mAh 3S battery and the 5000 mAh 6S battery consistently outperform the smaller battery configurations. Although larger batteries reduce transportation multiplicity, the additional energy contained in each delivered battery more than compensates for this loss.

Another important observation is that the performance gap between battery types increases with mission distance. At short distances, all battery types perform reasonably well because relay amplification remains limited. However, at larger distances, the relay process increasingly favors batteries with higher energy content. The results therefore suggest that battery selection becomes more critical as the relay chain grows deeper.

\subsection{Impact of Porter-Drone Capability}

To evaluate the influence of drone characteristics, we compare the light, medium, and heavy porter-drone classes defined previously. For this experiment, the 5000 mAh 3S battery is used as the relay battery, while the station inventory remains fixed at 20 porter drones and 100 batteries.

The corresponding transportation capacities are approximately 1, 6, and 24 batteries per flight for the light, medium, and heavy porter drones, respectively. These large differences in carrying capacity directly affect the relay amplification process.

\begin{figure}[t]
\centering
\includegraphics[width=0.9\linewidth]{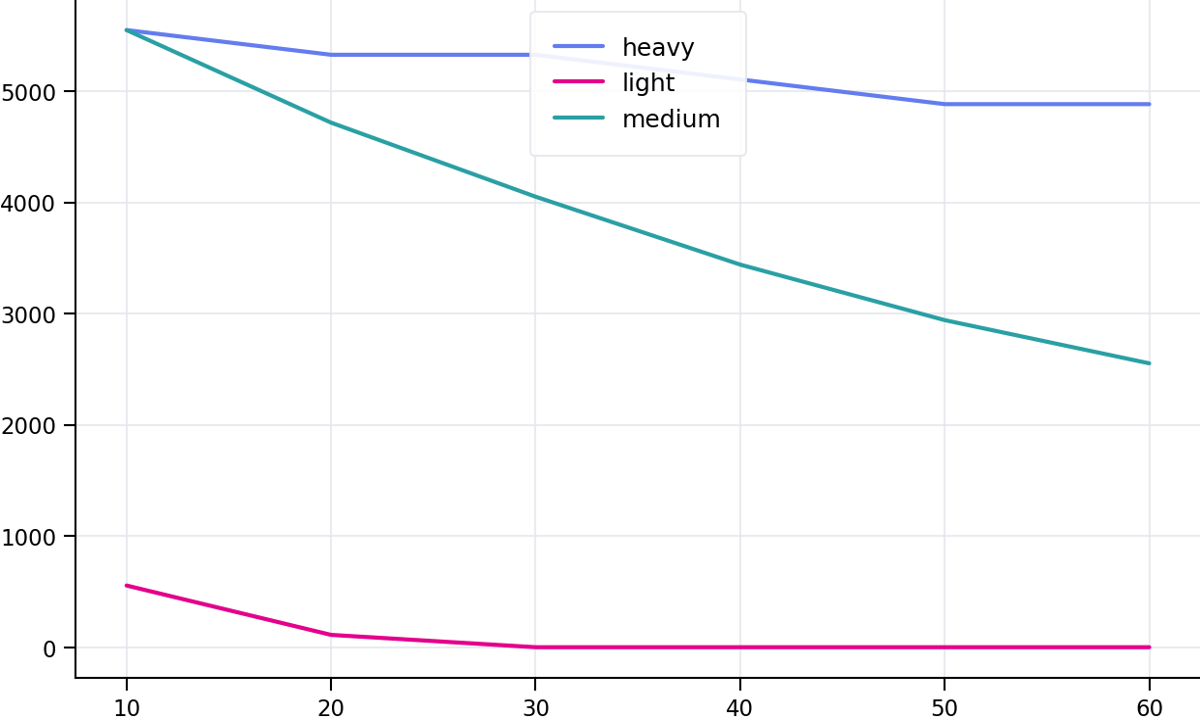}
\caption{Deliverable energy versus mission distance for different porter-drone classes using a 5000 mAh 3S battery.}
\label{fig:drone_comparison}
\end{figure}

Figure~\ref{fig:drone_comparison} shows that the heavy porter drone consistently achieves the highest deliverable energy, followed by the medium porter drone and then the light porter drone. The primary reason is the payload-dependent transportation capacity $K_{d,b}$. A larger payload allows more relay batteries to be transported per flight, thereby reducing relay amplification and improving delivery efficiency.

In addition, larger drone platforms typically provide longer relay ranges, which reduce the required relay depth. The combined effect of larger transportation capacity and lower relay depth substantially improves energy-projection efficiency.

Another notable observation is that the performance gap among drone classes increases with mission distance. For short-distance missions, all drone classes can deliver a meaningful amount of energy. As relay depth increases, however, the limited payload capacity and range of light porter drones result in a rapid reduction in deliverable energy. In contrast, heavy-lift drones retain a considerably larger fraction of their delivery capability.

These results suggest that payload capacity is a dominant factor in deep ILR operations. Although heavy drones may incur higher acquisition and operating costs, they can significantly reduce relay overhead and improve energy-projection efficiency.


First, battery selection must be considered jointly with drone payload capacity. The relay process introduces integer transportation constraints that make battery choice far more important than a simple comparison based on energy density alone.
Second, porter-drone capability has a strong impact on long-distance energy projection. Increasing payload capacity and relay range simultaneously improves transportation capacity and reduces relay depth, thereby mitigating relay amplification.


\section{Conclusion}
We introduced a new problem formulation for sustaining UAV missions in infrastructure-less regions (ILRs), where conventional charging infrastructure and ground-supported energy logistics are unavailable, and proposed a Distributed Energy Relay Framework and a Transportation-by-Relay-and-Cache (TRC) mechanism that enables cooperative drones to transport battery resources to distant mission locations while guaranteeing their own safe return through distributed battery caching.
Building on this, we developed models for distributed stations external to the ILR to compute energy-delivery capability, delivery cost, and reliability, and bid for energy provisioning. Distributed stations can, thus, cooperatively satisfy mission energy requirements while minimizing delivery cost.

Our vision for UAV systems extends beyond energy delivery to the broader Drone-as-a-Service (DaaS) ecosystem envisioned in OmniDroneX \cite{yen2026omnidronex}. Future work will further investigate energy provisioning and relay mechanisms and incorporate them as supporting services within this holistic UAV platform.

\bibliographystyle{IEEEtran} 
\bibliography{energy-relay}

\end{document}